\documentclass[aps,prl,superscriptaddress,twocolumn,nopacs,amsmath,amssymb,letter]{revtex4-1}

\setcitestyle{super}
\usepackage{color}
\usepackage{graphicx}% Include figure files
\usepackage{dcolumn}% Align table columns on decimal point
\usepackage{bm}% bold math
\usepackage{ulem}
\usepackage{xspace}

\def \EI {$E_\mathrm{ISB}$\xspace}

\begin{document}

\title{Maximal Rashba-like spin splitting via kinetic energy-driven\\ inversion symmetry breaking}

\author{Veronika Sunko}
\affiliation {SUPA, School of Physics and Astronomy, University of St. Andrews, St. Andrews KY16 9SS, United Kingdom}
\affiliation {Max Planck Institute for Chemical Physics of Solids, N{\"o}thnitzer Stra{\ss}e 40, 01187 Dresden, Germany}

\author{H.~Rosner}
\author{P.~Kushwaha}
\author{S.~Khim}
\affiliation {Max Planck Institute for Chemical Physics of Solids, N{\"o}thnitzer Stra{\ss}e 40, 01187 Dresden, Germany}

\author{F.~Mazzola}
\author{L.~Bawden}
\author{O.~J.~Clark}
\affiliation {SUPA, School of Physics and Astronomy, University of St. Andrews, St. Andrews KY16 9SS, United Kingdom}

\author{J.~M.~Riley}
\affiliation {SUPA, School of Physics and Astronomy, University of St. Andrews, St. Andrews KY16 9SS, United Kingdom}
\affiliation{Diamond Light Source, Harwell Campus, Didcot, OX11 0DE, United Kingdom}

\author{D.~Kasinathan}
\author{M.~W.~Haverkort}
\affiliation {Max Planck Institute for Chemical Physics of Solids, N{\"o}thnitzer Stra{\ss}e 40, 01187 Dresden, Germany}

\author{T.~K.~Kim}
\author{M.~Hoesch}
\affiliation{Diamond Light Source, Harwell Campus, Didcot, OX11 0DE, United Kingdom}

\author{J.~Fujii}
\author{I.~Vobornik}
\affiliation{Istituto Officina dei Materiali (IOM)-CNR, Laboratorio TASC, in Area Science Park, S.S.14, Km 163.5, IW34149 Trieste, Italy}

\author{A.~P.~Mackenzie}
\email{Andy.Mackenzie@cpfs.mpg.de}
\affiliation {SUPA, School of Physics and Astronomy, University of St. Andrews, St. Andrews KY16 9SS, United Kingdom}
\affiliation {Max Planck Institute for Chemical Physics of Solids, N{\"o}thnitzer Stra{\ss}e 40, 01187 Dresden, Germany}

\author{P.~D.~C.~King}
\email{philip.king@st-andrews.ac.uk}
\affiliation {SUPA, School of Physics and Astronomy, University of St. Andrews, St. Andrews KY16 9SS, United Kingdom}

\date{\today}% It is always \today, today,
             
\maketitle 

{\bf Engineering and enhancing inversion symmetry breaking in solids is a major goal in condensed matter physics and materials science, as a route to advancing new physics and applications ranging from improved ferroelectrics for memory devices to materials hosting Majorana zero modes for quantum computing. Here, we uncover a new mechanism for realising a much larger energy scale of inversion symmetry breaking at surfaces and interfaces than is typically achieved. The key ingredient is a pronounced asymmetry of surface hopping energies, i.e. a kinetic energy-driven inversion symmetry breaking, whose energy scale is pinned at a significant fraction of the bandwidth. We show, from spin- and angle-resolved photoemission, how this enables surface states of $3d$ and $4d$-based transition-metal oxides to surprisingly develop some of the largest Rashba-like spin splittings that are known. Our findings open new possibilities to produce spin textured states in oxides which exploit the full potential of the bare atomic spin-orbit coupling, raising exciting prospects for oxide spintronics. More generally, the core structural building blocks which enable this are common to numerous materials, providing the prospect of enhanced inversion symmetry breaking at judiciously-chosen surfaces of a plethora of compounds, and suggesting routes to interfacial control of inversion symmetry breaking in designer heterostructures.
}

The lifting of inversion symmetry is a key prerequisite for stabilising a wide range of striking physical properties such as chiral magnetism, ferroelectricity, odd-parity multipolar orders, and the creation of Weyl fermions and other spin-split electronic states without magnetism ~\cite{nagaosa_topological_2013,dupe_engineering_2016, lee_strong_2005,lee_strong_2010,fu_parity-breaking_2015, xu_discovery_2015,mourik_signatures_2012}. Inversion symmetry is naturally broken at surfaces and interfaces of materials, opening exciting routes to stabilise electronic structures distinct from those of the bulk~\cite{hwang_emergent_2012,mannhart_oxide_2010,manchon_new_2015,hasan_colloquium:_2010,zubko_interface_2011,xiao_interface_2011,caviglia_tunable_2010,ast_giant_2007}. A striking example is found in materials which also host significant spin-orbit interactions, where inversion symmetry breaking (ISB) underpins the formation of topologically-protected surface states~\cite{hsieh_topological_2008,hasan_colloquium:_2010} and Rashba~\cite{bychkov_properties_1984,manchon_new_2015} spin splitting of surface or interface-localised two-dimensional electron gases~\cite{lashell_spin_1996,nitta_gate_1997,hoesch_spin_2004,ast_giant_2007,caviglia_tunable_2010,king_large_2011,el-kareh_quantum_2013}, generically characterised by a locking of the quasiparticle spin perpendicular to its momentum. Such effects lie at the heart of a variety of proposed applications in spin-based electronics~\cite{datta_electronic_1990,koo_control_2009,manchon_new_2015,mellnik_spin-transfer_2014,lesne_highly_2016,wunderlich_spin_2010}, and provide new routes to stabilise novel physical regimes such as spiral RKKY interactions, enhanced electron-phonon coupling, localisation by a weak potential, large spin-transfer torques, and mixed singlet-triplet superconductivity \cite{chesi_rkky_2010, lai_spatial_2009,cappelluti_electron-phonon_2007, grimaldi_weak-and_2008, chaplik_bound_2006, gorkov_superconducting_2001,nam_ultrathin_2016, matetskiy_two-dimensional_2015}.
Conventional wisdom about how to maximize the Rashba effect has been to work with heavy elements whose atomic spin-orbit coupling is large.  However, the energetic spin splittings obtained are usually only a small fraction of the atomic spin-orbit energy scale.  This is because the key physics is not exclusively that of spin-orbit coupling, but rather an interplay between the spin-orbit, $E_\mathrm{SOC}$, and inversion breaking, \EI, energy scales. 

%FIGURE 1
\begin{figure}
\includegraphics[width=0.9 \columnwidth]{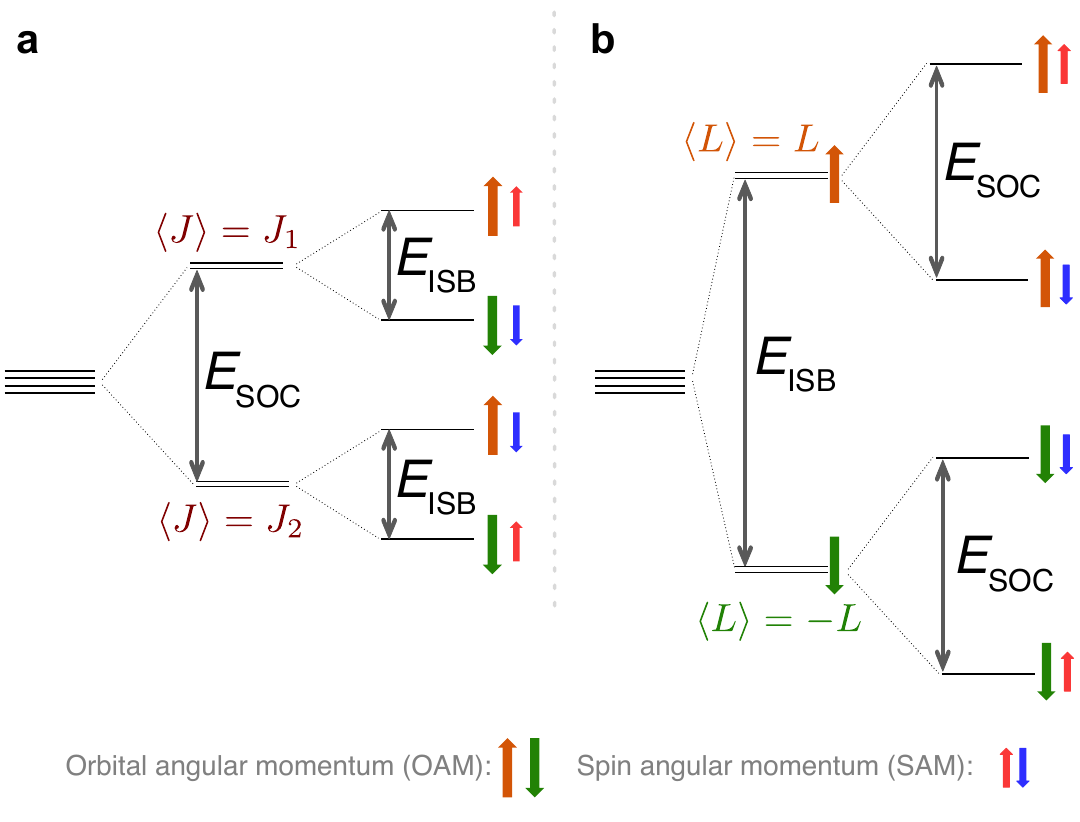}
\caption{ \label{f:overview} {\bf Interplay of two energy scales for realising spin splitting in inversion asymmetric environments.} Schematic illustration of how a four-fold degenerate state becomes split by ISB and SOC. (a) A dominant SOC first forms states of defined total angular momentum, $J$, split by the atomic SOC ($E_{\mathrm{SOC}}$) and where the orbital (OAM) and spin (SAM) angular momentum are locked to each other. ISB splits them additionally into states of opposite spin, separated by \EI. (b) If \EI dominates, the initial splitting is between states of opposite OAM. The weaker SOC cannot mix the states split by ISB, and thus introduces splitting between the states of spin parallel and antiparallel to the preexisting OAM, with a magnitude $\sim\!E_{\mathrm{SOC}}$. For a given SOC strength, the spin splitting is therefore largest if $E_{\mathrm{ISB}}\gg{E}_\mathrm{SOC}$.}
\end{figure}

%FIGURE 2
\begin{figure*} [!t]
\includegraphics[width=\textwidth]{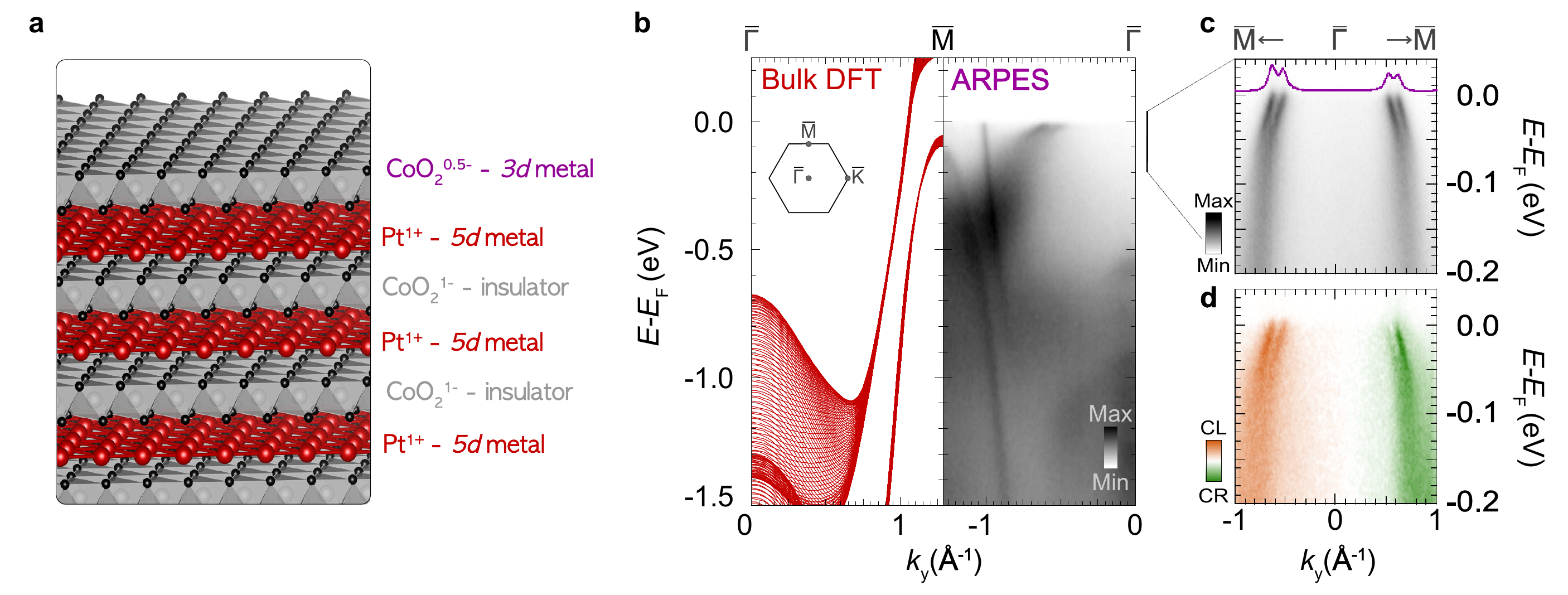}
\caption{ \label{f:PtCoO2_summary} {\bf Bulk and surface electronic structure of PtCoO$_2$.} (a) The layered PtCoO$_2$ delafossite structure, with linearly-coordinated metallic Pt (formally $1+$) layers separated by (CoO$_2$)$^{1-}$. Such ionic pictures suggest a metallic (CoO$_{2}$)$^{0.5-}$ surface layer. (b) Both $k_z$-projected bulk density functional theory (DFT) calculations and an ARPES measurement ($h\nu=~$110 eV, $s$-polarised light) of the electronic structure along the $\overline{\Gamma}-\overline{M}$ direction of the surface Brillouin zone reveal a 2D band crossing the Fermi level and a fully occupied hole band centred at the $\overline{M}$ point. Two additional hole-like bands crossing the Fermi level are evident in (b) and shown magnified in (c) for measurement conditions ($h\nu=~$110 eV, $p$-pol) which increase their spectral weight. We attribute these to surface states of the top CoO$_{2}$ layer. They host ``kinks'' close to the Fermi level, as well as increased broadening away from $E_F$, which are characteristic spectroscopic signatures of many-body interactions. Momentum distribution curves (MDCs) at the Fermi level ($E_{F}\pm5~meV$) are shown by the purple lines. (d) Experimental circular dichroism measurements (h$\nu$ =110~eV) suggest that these surface states host a strong OAM, of the same sign for each surface state branch.}
\end{figure*}

Evidencing the importance of this interplay, the size of the spin splitting can often be controlled by tuning the strength of the inversion symmetry-breaking potential, for example via applied electrostatic gate voltages ~\cite{nitta_gate_1997}. This indicates a regime where the achievable spin-splitting is, in fact, limited by \EI, typically to only a modest percentage of the atomic SOC (Fig. ~\ref{f:overview}(a)) \cite{hong_quantitative_2015, park_microscopic_2015, park_orbital-angular-momentum_2011}. If, however, an ISB energy scale can be achieved which is significantly larger than the SOC, the spin-orbit Hamiltonian will act as a perturbation to a Hamiltonian dominated by the ISB. A four-fold degenerate level will be split by ISB into two states of opposite orbital angular momentum (OAM). The weaker spin-orbit interaction cannot mix those states, instead further spin-splitting them into states of spin parallel and antiparallel to the preexisting OAM, with a splitting that must therefore take the full atomic SOC value (Fig.~\ref{f:overview}(b))\cite{hong_quantitative_2015, park_microscopic_2015, park_orbital-angular-momentum_2011}. This is clearly a desirable limit to maximise the achievable spin-splitting for a given strength of SOC, but is typically only realised when the absolute value of SOC, and consequently also the total spin splitting, is small.

Here, we demonstrate a new route which exploits the intrinsic energetics of hopping to realise large inversion symmetry breaking energy scales at the surfaces of transition-metal-based delafossite oxides (Pd,Pt)(Co,Rh)O$_2$. We show, from spin- and angle-resolved photoemission (ARPES) how this leads to spin-split Fermi surfaces  hosting one of the largest momentum separations known even for the unlikely environment of a 3d-electron based CoO$_2$ metal. Crucially, we demonstrate that \EI scales as the bandwidth, and so will itself grow concomitant with the SOC strength upon moving to heavier-element systems. We show how this allows the full atomic SOC to be retained in spin-splitting the states of a $4d$ RhO$_2$ surface layer, leading to an unprecedented spin splitting for an oxide compound of $\sim\!150$~meV, and opening entirely new opportunities to tune between truely giant spin splittings and strongly-interacting Rashba-like states in oxides and other compounds.

We consider first the delafossite oxide~\cite{shannon_chemistry_1971,shannon_chemistry_1971-1,shannon_chemistry_1971-2,tanaka_growth_1996,takatsu_roles_2007,mackenzie_properties_2017} PtCoO$_{2}$. Pt$^{1+}$ cations sit on a triangular lattice, leading to extremely high bulk conductivity~\cite{kushwaha_nearly_2015}, and are separated by layers of corner-sharing and trigonally-distorted cobalt oxide octahedra (Fig.  ~\ref{f:PtCoO2_summary}(a)). In the bulk, six $3d$ electrons fill the Co $t_{2g}$ manifold, rendering the CoO$_{2}$ block insulating~\cite{kushwaha_nearly_2015}. At the surface, however, the `missing' Pt atoms from above the topmost oxygen layer will cause the CoO$_2$ block to have a formal valence of only $0.5^-$, effectively hole-doping the manifold of Co-derived states~\cite{kim_fermi_2009}. Consistent with this simple ionic picture, our angle resolved photoemission (ARPES) data not only show sharp and rapidly dispersing spectral features representative of the bulk band structure, but also a pair of two-dimensional hole bands crossing the Fermi level where none are expected from the calculated bulk electronic structure at any value of k$_{z}$  (Figs.~\ref{f:PtCoO2_summary}(b) and Extended Data Fig.~1). 

%FIGURE 3
\begin{figure*}[!t]
\begin{center}
\includegraphics[width=0.7\textwidth]{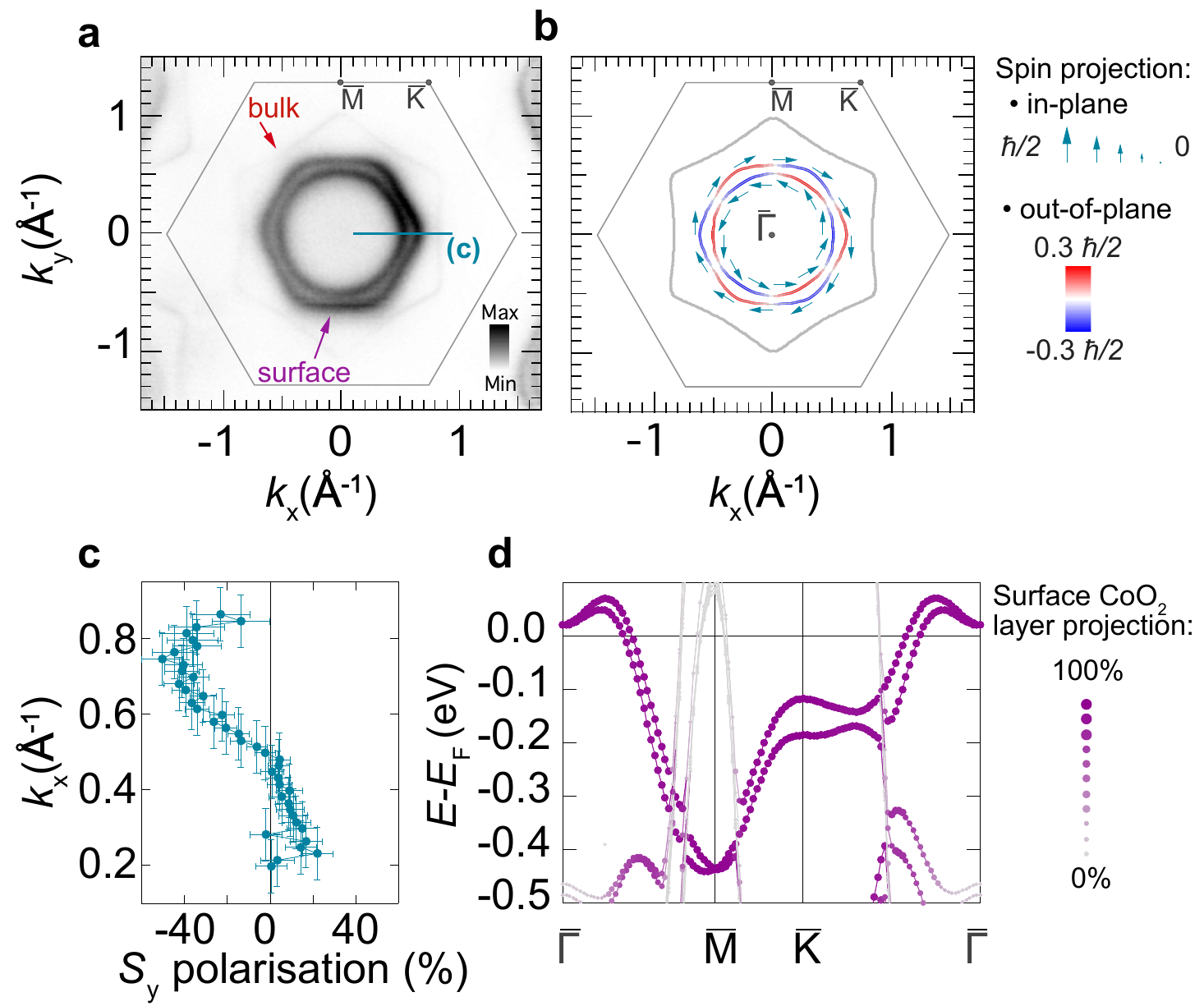}
\caption{ \label{f:spin} {\bf Spin-polarised surface states.} Bulk and surface Fermi surfaces of PtCoO$_{2}$ (a) measured by ARPES (E$_{F}\pm$5~meV, $h\nu=$110~eV, $p$-pol) and (b) from a CoO$_2$-terminated DFT supercell calculation. Spurious bulk hole pockets near $\overline{M}$ from the calculations are not shown (see Methods). (c) Spin-ARPES measurements ($h\nu=$65~eV, $p$-pol) of an in-plane spin polarisation ($\langle{S_y}\rangle$) of the Fermi surface for the cut along $k_{x}$ shown in (a), revealing a Rashba-like spin texture of the surface states. This is supported by our supercell calculations (b) which show a chiral in-plane spin texture as well as some out-of-plane spin canting.  The length of the arrows encodes the modulus of the expectation value of spin along a quantization axis perpendicular to the in-plane momentum; the expectation value parallel to the momentum direction is zero. $\langle{S_z}\rangle$ is shown by the colour. (d) Electronic structure from the DFT supercell calculations projected onto the first CoO$_2$ layer, demonstrating how the bands that make up these Fermi surfaces have almost all of their wavefunction weight on the surface CoO$_2$ block. }
\end{center}
\end{figure*}

The inner and outer band form almost circular and hexagonally-warped hole-like Fermi surfaces, respectively, located at the Brillouin zone centre (Fig.~\ref{f:spin}(a)). They host heavy quasiparticle masses of up to 15$m_e$, where $m_e$ is the free-electron mass (see Extended Data Table~S1 for a more detailed description of the quasiparticle masses and their anisotropy), as well as clear spectroscopic signatures of both electron-phonon and electron-electron interactions (Fig.\ref{f:PtCoO2_summary}(b, c)). This is in stark contrast to the almost free-electron masses of the bulk Pt-derived bands crossing $E_F$~\cite{kushwaha_nearly_2015}, and instead suggests that these states are predominantely derived from much more local Co $3d$ orbitals. Consistent with this, we note that similar dispersions have been observed previously in the sister compound PdCoO$_{2}$~\cite{noh_anisotropic_2009}. On this basis, as well as our density-functional calculations discussed below, we thus attribute these as surface states of the CoO$_2$-terminated surface of PtCoO$_2$, with very little intermixing of Pt.

Intriguingly, our ARPES measurements performed using circularly-left and circularly-right polarised light reveal a strong circular dichroism (CD) of these surface states (Fig.~\ref{f:PtCoO2_summary}(d) and Extended Data Fig.~1). CD-ARPES is known to be sensitive to OAM structures in solids~\cite{park_chiral_2012, park_orbital_2012}. While OAM would naturally be expected to be quenched for the $t_{2g}$ manifold, the observation of a pronounced circular dichroism across the entire bandwidth of the surface states here suggests that these instead host a large OAM. Moreover, from spin-resolved ARPES measurements (Fig.~\ref{f:spin}(c), see Methods) we find that the two surface states are strongly spin-polarised, with a significant in-plane spin component pointing perpendicular to their momentum. This indicates chiral spin textures of the form that would be expected for a Rashba-like splitting ~\cite{bychkov_properties_1984}. Our DFT calculations for a CoO$_2$-terminated supercell (Fig.~\ref{f:spin}(b)) reproduce such spin-momentum locking of the surface Fermi surfaces, additionally revealing a small out-of-plane spin canting. 

While the spin chirality reverses sign between the inner and outer Fermi surface sheets, the circular dichroism of each spin-split branch is of the same sign (Fig.~\ref{f:PtCoO2_summary}(d) and Extended Data Fig.~1). This suggests that both branches carry the same sign of chiral OAM, with spin splitting resulting from the parallel and anti-parallel alignment of the spin to this OAM in striking concord with the situation shown in Fig.~\ref{f:overview}(b). Critically, this should lead to a spin splitting which is on the order of the full strength of atomic SOC. Our DFT supercell calculations (Fig.~\ref{f:spin}(d)) fully support this assignment. As well as confirming that the surface state wavefunction is almost entirely located in the topmost CoO$_2$ block, they reveal that these cobalt-derived states host a spin splitting which remains large over their entire bandwidth, except in close proximity to time-reversal invariant momenta ($\overline{\Gamma}$ and $\overline{M}$) where time-reversal symmetry dictates that it must vanish. Away from these points, it reaches as high as 60~meV at the average $k_{F}$ along the $\overline{\Gamma}-\overline{K}$ direction, which is comparable to the atomic SOC strength of Co~\cite{haverkort_spin_2005}.

%FIGURE 4
\begin{figure}
\begin{center}
\includegraphics[width= 0.9 \columnwidth]{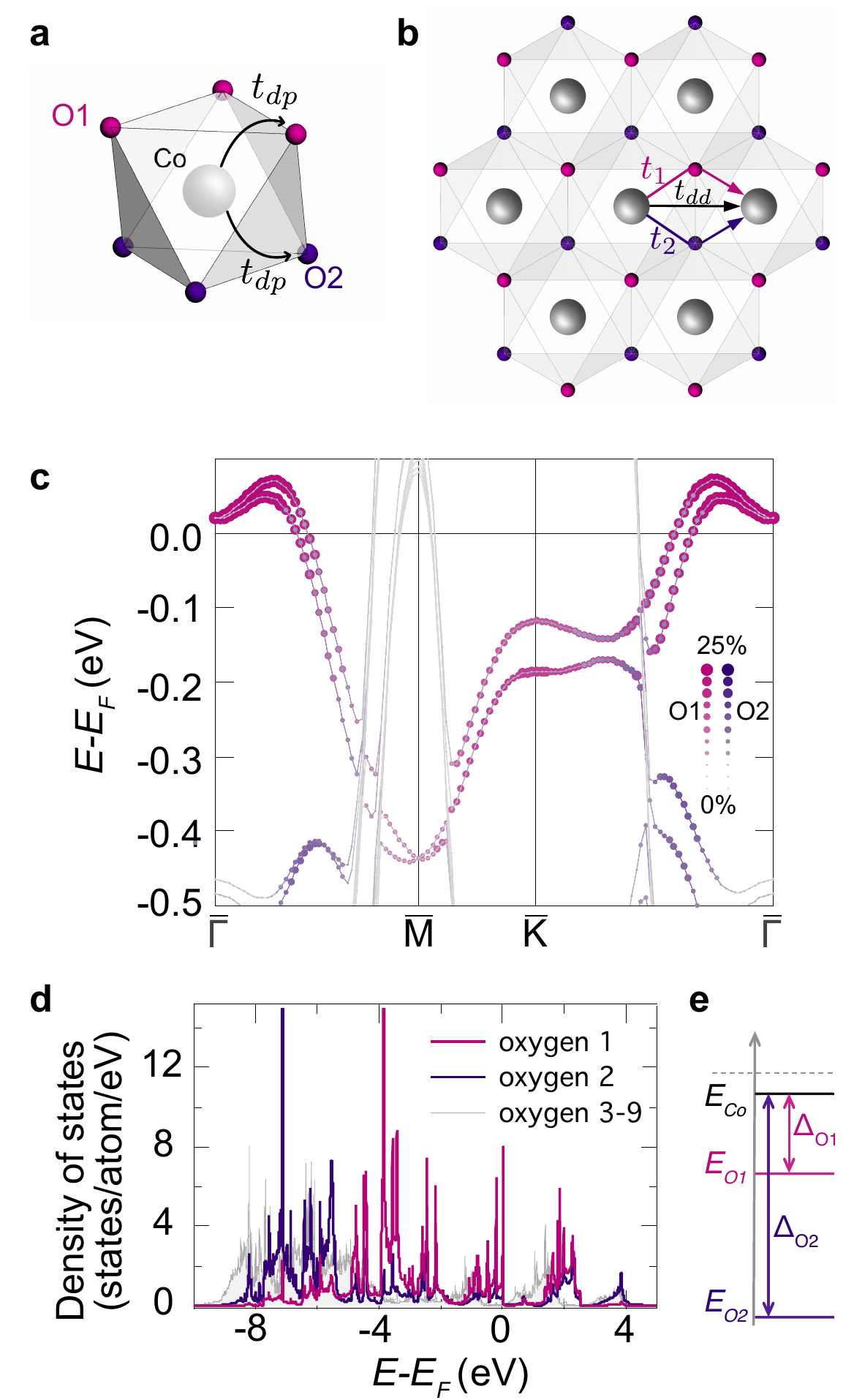}
\caption{ \label{f:DFT} {\bf Origin of the spin splitting.} (a) A single Co-O octahedron. The hopping integral between Co and the oxygen in layers above (O1, pink) and below (O2, purple) the Co is denoted $t_{dp}$. (b) Top view of the surface CoO$_{2}$ layer. As shown schematically, an electron can hop between two Co atoms either directly (t$_{dd}$), or via the oxygen atoms (effective hopping integrals $t_{1}$ and $t_{2}$, for O1 and O2, respectively). (c) DFT supercell  band structure calculations, projected onto O1 (pink) and O2 (purple), revealing the dominant contribution to the surface states is from O1, the oxygen layer above Co. (d) The layer-resolved oxygen partial density of states (PDOS) indicates how this arises due to a $\sim\!4$~eV shift to lower binding energy of the PDOS of O1 with respect to that of O2, represented schematically as an on-site energy shift in (e). The PDOS of O2 is similar to that of the bulk-like oxygens in deeper layers (O3-O9, grey lines in (d)).}
\end{center}
\end{figure}

Spin-orbit interactions are generally neglected for $3d$-orbital systems. In contrast, our work demonstrates how they can lead to a major restructuring of the electronic structure if a sufficiently strong ISB can be realised to unlock their full strength, a point we return to below. This opens exciting new potential to investigate the interplay of spin-orbit interactions with strong electronic correlations - something which is more naturally associated with local $3d$ orbitals~\cite{imada_metal-insulator_1998,dagotto_complexity_2005}. Already here, achieving the maximal possible energetic spin splitting combined with high quasiparticle masses leads to momentum spin splittings at the Fermi level (the relevant quantity for a number of technological applications) as high as $\Delta{k_F}=(0.13\pm 0.01)$~\AA$^{-1}$ (see also Extended Data Table 1). This is amongst the highest known for any Rashba-like system~\cite{ishizaka_giant_2011}. It is approximately ten times larger than that of the enhanced Rashba-like splitting thought to occur at isolated momentum points where different $t_{2g}$ bands intersect in the $3d$ orbital system of SrTiO$_3$-based two-dimensional electron gases~\cite{caviglia_tunable_2010,king_quasiparticle_2014,zhong_theory_2013,khalsa_theory_2013}, and is in fact comparable in magnitude to the momentum splitting in the so-called giant Rashba semiconductor BiTeI~\cite{ishizaka_giant_2011,bawden_hierarchical_2015}, which has both very strong SOC and large internal electrostatic potential gradients.  We note that the experimentally observed Fermi surfaces are well reproduced by our DFT calculations which consider an ideal bulk-truncated supercell. This rules out subtle surface structural modulations, such as those known to dramatically enhance Rashba spin splitting in various surface alloy systems~\cite{ast_giant_2007,bihlmayer_enhanced_2007,bian_origin_2013}, as the origin of the large effects observed here. Instead, it identifies the spin splitting which we observe as an intrinsic property of the bulk-like CoO$_2$ layer which is unlocked when it is placed in an environment where inversion symmetry is broken by the presence of the surface.

We now show how this occurs as a natural consequence of a strong asymmetry in effective Co-Co hopping paths through oxygen atoms located above and below the topmost Co layer. The structure of a single CoO$_2$ layer is shown in Fig.~\ref{f:DFT}(a,b). This can be viewed as a tri-layer unit, with a central triangular net of Co atoms, above and below which are two triangular oxygen sublattices which have opposite orientation with respect to Co. Given the local nature of $3d$ orbitals, a key Co-Co effective hopping will be via oxygen, either through the oxygen layer above (O1) or below (O2) the transition-metal plane. In bulk, these must be equivalent due to the inversion symmetry of the crystal structure. This requirement is, however, lifted at the surface. To demonstrate this explicitly, we show in Fig.~\ref{f:DFT}(c) the projection of the electronic structure from our DFT supercell calculations onto O1 and O2 of the surface CoO$_2$ layer. It is clear that there is a significantly higher admixture of O1 than O2 in the CoO$_2$-derived surface states which intersect the Fermi level identified above: we estimate that the O1 orbital contribution is approximately 2.4 times higher than that of O2 when averaged around the surface Fermi surfaces. This indicates that Co-O-Co hopping occurs dominantly via the oxygen network above, rather than below, the transition-metal plane in the surface CoO$_2$ layer.

The microscopic origin of this is evident from the layer-projected oxygen partial density of states (PDOS) shown in Fig.~\ref{f:DFT}(d). While the PDOS of O2 is similar to those of subsequent layers of our supercell calculation, that of O1 is strongly shifted in energy, with the main centroid of spectral weight located at $\sim\!4$~eV lower binding energy compared to that of O2 (see also Extended Data Fig.~2). In the bulk~\cite{kushwaha_nearly_2015} and in the bulk-like environment of O2, the bonding Pt-O combinations shift the predominantly O-derived levels to higher binding energy. An absence of this shift due to the missing Pt above O1 therefore causes the difference in on-site energy of the two oxygens. In a tight-binding picture, Co-O-Co hopping via O$n$ ($n=1,2$) is described by the effective transfer integral, $t_n=t_{dp}^2/(E_{Co}-E_{On})$ where $t_{dp}$ is the Co-O hopping matrix element (Fig.~\ref{f:DFT}(a)) and $E_{Co}$ ($E_n$) is the on-site energy of Co (O$n$) (Fig.~\ref{f:DFT}(e)). The decrease in binding energy of the oxygen-derived PDOS on site 1 leads to a strongly asymmetric hopping with $t_1>t_2$, and thus directly to a large \EI. 

To validate that this drives the spin splitting observed here, we develop a minimal tight-binding model for a single CoO$_2$ layer (see Methods and Extended Data Fig.~3). This yields an estimate of the relative ISB introduced by the asymmetric hopping of $\alpha_\mathrm{ISB}=(t_{1}-t_{2})/(t_{1}+t_{2})>40 \%$. This asymmetry enters between two hopping paths, nominally equivalent in bulk, which define the entire bandwidth, and thus the energy scale of the ISB is large: $E_\mathrm{ISB}=\alpha_\mathrm{ISB}t\approx 150$~meV, where $t$ is the average effective Co-O-Co hopping integral. This is approximately twice the atomic SOC, and so places the CoO$_2$ surface within the ``strong ISB'' limit discussed above (Fig.~\ref{f:overview}(b)), explaining why the spin splitting is so large for the Co-derived states observed here.

Critically, the analysis presented here shows how the inversion symmetry breaking can enter directly via the dominant kinetic term of the Hamiltonian. This differs substantially from previous treatments, which typically consider the symmetry breaking as arising directly from the surface electric field~\cite{hong_quantitative_2015}. There, the relevant energy scale is the dipole energy, which is only be a weak perturbation to a Hamiltonian governed by the kinetic energy~\cite{khalsa_theory_2013}. In contrast, the ISB achieved here is proportional to the bandwidth. This suggests exciting new strategies to maximise spin-splitting in solids which are counter to conventional wisdom: instead of just increasing the SOC, which would normally lead to the achievable spin splitting becoming limited by \EI (Fig.~\ref{f:overview}(a)), the ``bandwidth-scaled''  \EI will simultaneously grow with increased orbital overlap of heavy-element systems, enabling the maximum possible spin splittings to be retained even in much stronger SOC systems.

To verify the validity and potential of this new approach, it is desirable first to experimentally compare the surface states of PtCoO$_2$ with those of PdCoO$_2$, to rule out an influence of the strong SOC of the noble-metal cation in mediating our observed spin splittings, and then to change the transition metal from Co to Rh, to test whether the spin splitting really scales with the factor of  $\sim\!2.5$ increase of transition-metal SOC.  The former is easy to do, since high quality PdCoO$_2$ crystals are straightforward to synthesize\cite{takatsu_roles_2007, hicks_quantum_2012}.  However, PdRhO$_2$, though known to be metallic in polycrystalline form\cite{shannon_chemistry_1971-1}, has not to our knowledge previously been synthesized in single-crystal form.  We did so for this project.  

\begin{figure}
\begin{center}
\includegraphics[width= 0.9 \columnwidth]{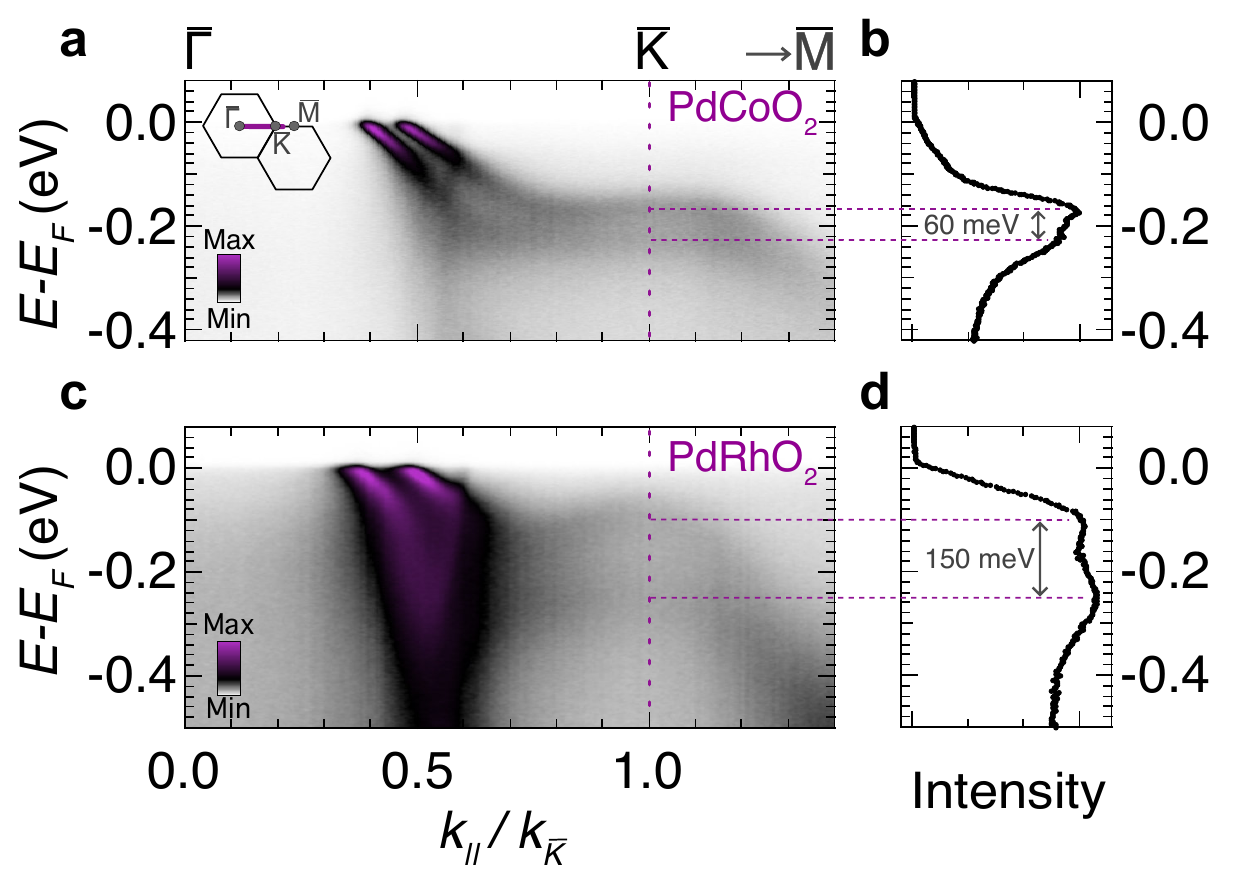}
\caption{ \label{f:CoRh} {\bf Bandwidth-scaled inversion symmetry breaking.} Spin-split surface states of (a) PdCoO$_2$ and (c) PdRhO$_2$, indicating a spin splitting which retains the strength of the atomic SOC of the transition metal even as this is increased by a factor of approximately 2.5 on moving from Co to Rh. This is particularly evident from the spin splitting at the ${\overline{K}}$-point, clearly visible in EDCs (b,d). }
\end{center}
\end{figure}

As shown in Fig.~\ref{f:CoRh}(a) and Extended Data Table 1, the surface state spin-splitting of PdCoO$_2$ is in good quantitative agreement with that of PtCoO$_2$. This further confirms that the spin splitting we observe is a property of the CoO$_2$ block, with only minimal influence from the A-site (Pd,Pt) cation. We can clearly resolve the spin-split bands across their bandwidth,  extracting a spin splitting of  60~meV at the $\overline{K}$-point  (Fig.~\ref{f:CoRh}(b)) . As shown in our tight-binding model (Extended Data Fig.~3(a)), a four-fold degenerate band crossing would be expected at this momentum point in the absence of ISB and SOC. It is therefore an ideal location to probe the energy scale of the spin splitting. 

%FIGURE 5

Our measurements of PdRhO$_2$ (Fig.~\ref{f:CoRh}(c)) reveal qualitatively similar surface states. The momentum splitting at the Fermi level is, however, higher, reaching $\Delta{k_F}=(0.16\pm0.01)$~\AA$^{-1}$ along $\overline{\Gamma}-\overline{K}$, despite its lower quasiparticle masses (see also Extended Data Table 1). This is a result of a strongly enhanced energetic splitting, which we find is  150~meV at the $\overline{K}$-point (Fig.~\ref{f:CoRh}(d)). To the best of our knowledge, this is by far the largest spin splitting observed in an oxide to date, paving the way for new applications in oxide spintronics. More importantly, it is on the order of the full atomic SOC strength of Rh$^{3+}$ (Ref.~\cite{haverkort_spin_2005}). The bandwidth scaling of \EI introduced above allows PdRhO$_{2}$ to remain in a limit where $E_{\mathrm{ISB}}>E_{\mathrm{SOC}}$ even as the latter is markedly increased. This is further justified by our CD-ARPES measurements (Extended Data Fig.~4), where both of the spin-split branches retain the same sign of chiral OAM, just as in (Pt,Pd)CoO$_2$. 

This key finding should not be specific to the delafossite compounds studied here. Rather it should be generally applicable to systems where hopping between the frontier orbitals is mediated by atoms above and below their plane, such as, for example, at the $(111)$-oriented surfaces and interfaces of perovskites. Large on-site energy shifts of the surface layer, naturally expected due to the loss of bonding and subject to further manipulation by, e.g., applied electric or strain fields, directly introduce a large asymmetry in otherwise equivalent hopping paths in such structures, and could thus generically be expected to lead to large spin-orbit induced spin splittings. Immediate quantitative insight into the levels of inversion symmetry breaking expected could in principle be obtained from density functional theory calculations. This is therefore an example of reliable predictive capability of the "materials by design" approach that is currently generating large interest from specialists in first-principles electronic structure calculations. Moreover, suitable mismatched interfaces could be fabricated by highly-controlled thin film and multilayer techniques such as molecular beam epitaxy, opening exciting possibilities to design and access novel physical regimes.

\

\noindent{\bf Methods}\\
{\small 
\noindent{\bf ARPES:} Single-crystal samples of PtCoO$_2$, PdCoO$_2$, and PdRhO$_2$ were grown by flux and vapour transport techniques in sealed quartz tubes. They were cleaved {\it in-situ} at the measurement temperature of $\sim\!6-10$~K. Both (Pt,Pd)- and (Co,Rh)O$_2$-terminated surfaces would be expected for a typical cleaved surface. A-site terminated surface states have been reported previously for the sister compound PdCrO$_{2}$~\cite{sobota_electronic_2013}, and for some cleaves we observe states which we tentatively assign as derived from Pt/Pd terminations. However, in agreement with prior studies~\cite{noh_anisotropic_2009}, our ARPES spectra often show spectroscopic signatures arising only from the bulk and the CoO$_2$-terminated surface, which is the case for all of the data included here. ARPES measurements were performing using the I05 beamline of Diamond Light Source, UK. Measurements were performed using $s$-, $p$- and circularly-polarised synchrotron light from 55 to 110 eV and employing a Scienta R4000 hemispherical electron analyser. Spin-resolved ARPES measurements were performed using $p$-polarised synchrotron light at the APE beamline of the Elettra synchrotron~\cite{bigi_very_2016}, using a Scienta DA30 hemispherical analyser equipped with a very low-energy electron diffraction (V-LEED)-based spin polarimeter probing the spin polarisation perpendicular to the analyser slit, $\langle{S_y}\rangle$. The finite spin-detection efficiency was corrected using a Sherman function, S=0.3, determined by comparison with the known spin-polarisation of the Rashba-split surface states measured on the Au$(111)$ surface. The spin polarisation along MDCs at the Fermi level is extracted as 
\[P=\frac{I^+ - I^-}{S(I^+ + I^-)}\]
where $I^+$ ($I^-$) is the V-LEED channeltron intensity measured along MDCs for the target magnetisation polarised in the positive (negative) directions, defined with respect to the DFT calculation.

\

\noindent{\bf Density functional theory:} Relativistic density functional (DFT) electronic structure calculations were performed using the full-potential FPLO code~\cite{koepernik_full-potential_1999,opahle_full-potential_1999,fplo},  version fplo14.00-47. The exchange correlation potential was treated within the general gradient approximation (GGA), using the Perdew-Burke-Ernzerhof \cite{perdew_generalized_1996-1} parametrisation. Spin-orbit coupling (SOC) was treated non-perturbatively solving the four component Kohn-Sham-Dirac equation~\cite{eschrig_chapter_2004}. The influence of including spin-orbit coupling in the calculations, which leads to spin-splitting of the surface states, is shown explicitly in Extended Data Fig.~5. The density of states was calculated applying the tetrahedron method. For all calculations, the appropriate experimental crystal structures were used. The bulk electronic structure calculations were carried out on a well-converged mesh of 27000 $k$-points (30x30x30 mesh, 2496 points in the irreducible wedge of the Brillouin zone). The strong Coulomb repulsion in the Co-3d shell was taken into account in a mean field way, applying the GGA+$U$ approximation in the atomic-limit-flavour (AL), with U=4~eV~\cite{kushwaha_nearly_2015}.

The surface electronic structure was calculated using a symmetric slab containing 9 CoO$_{2}$ layers with Co in the centre, and separated by a vacuum gap of 15 \AA\ along the $z$ direction. The layer-projected Co partial density of states and the layer-resolved charge accumulation are shown in Extended Data Fig.~6. A dense $k$-mesh of 20x20x4 points was employed for the calculations. For this slab thickness the calculations are well converged with respect to the electronic states of the three central layers. However, due to the slight off-stoichiometry of the slab the partial charges  of these layers are  slightly modified from their true bulk values. We therefore set the Fermi level in the calculations to match the experimental crossing vectors of the surface states. This, however, causes a small upward shift of the hole-like bands near $\overline{M}$, causing these to intersect the Fermi level in the slab calculation (Fig.\ref{f:spin}(d)), unlike for the true bulk electronic structure (Fig.\ref{f:overview}(b)). We neglect these pockets in plotting the calculated Fermi surface (Fig.\ref{f:spin}(b)). Supercell calculations were also performed for fully-relaxed crystal structures, finding qualitatively the same results as in the calculations performed for the ideal truncated bulk crystals used for the results presented here.

\

\noindent{\bf Tight binding model:} To gain further understanding of the mechanism by which the difference in the on-site energy between the surface and subsurface oxygen stabilises a giant inversion symmetry breaking, ultimately leading to spin splitting of the full atomic spin-orbit coupling strength, we have developed a minimal tight binding model describing the surface CoO$_2$ layer (see Fig.~\ref{f:DFT}(a,b)). That is broadly justified by the DFT supercell calculations which indicate the surface states are almost entirely localised in the topmost CoO$_2$ layer. Since the model offers important insights into the key physics of this paper we describe it here at a level of detail that would allow it to be reproduced by any interested reader. 

A tight binding Hamiltonian was constructed using the Slater-Koster parametrisation of the energy integrals~\cite{slater_simplified_1954}, in the cubic $(xy, yz, 3z^2-r^2, xz, x^2-y^2, p1_{y}, p1_{z}, p1_{x}, p2_{y}, p2_{z}, p2_{x})$ basis, where $p1$ and $p2$ refer to the orbitals on the two distinct oxygens, and the $\hat{z}$ axis is taken to be normal to the crystal surface. For illustrative purposes and maximum simplicity, in Extended Data Fig.~3 we retain only Co-O-Co hopping between Co $t_{2g}$ and oxygen $p_z$ orbitals, which are dominant at the Fermi level, and neglect the small trigonal crystal field splitting of the $t_{2g}$ levels. We note that better agreement with our DFT calculations can be obtained by including both in-plane oxygen orbitals as well as direct Co $d$-$d$ hopping (Extended Data Fig.~7). 

Once the spin degree of freedom is taken into account the Hamiltonian is represented by a 22$\times$22 matrix. It is a sum of  three terms, the kinetic term $H_{K}$, the spin-orbit term $H_{SO}$, and the crystal field term  $H_{CF}$. The fact that the octahedral and crystallographic axes do not coincide (Fig.~\ref{f:DFT}(a)) requires a coordinate transformation between cubic and trigonal orbital basis sets. Here we discuss all of the contributions to the total Hamiltonian, as well as the free parameters of the model. 

The kinetic part of the Hamiltonian is constructed in the basis of cubic harmonics, with the $\hat{z}$ axis normal to the crystal surface. This choice of a coordinate system makes it easy to evaluate directional cosines between neighbouring atoms, $l=\sin{\theta}\cos{\phi}$, $m=\sin{\theta}\sin{\phi}$ and $n=\cos{\theta}$, where $\theta$ and $\phi$ are ~the usual polar angles.  The experimental crystal structure of PtCoO$_{2}$~\cite{kushwaha_nearly_2015}, with the hexagonal lattice constants of $a=2.82$\AA~ and  $c=17.8$\AA, and the Pt-O distance  $z=0.114~c$, was used to determine the relative positions of the atoms. If the Co atom is placed at the origin, $(0,0,0)$, the positions of two distinct oxygen atoms are given by $(\frac{1}{2},\frac{1}{2\sqrt[]{3}},0.33)~a$ and $(\frac{1}{2},-\frac{1}{2\sqrt[]{3}},-0.33)~a$. Extended Data Table~2 lists all of the hopping paths considered in the extended model, along with the angles between the nearest neighbour atoms  in the geometry outlined above, and the Slater-Koster (SK) parameters needed to describe the hopping~\cite{slater_simplified_1954}. 

The Co crystal field is diagonal in the trigonal basis, ($u_{+}$,$u_{-}$,$x_{0}$, $x_{1}$, $x_{2}$), where $u_{+}$ and $u_{-}$ are the two $e_{g}^{\sigma}$ orbitals, $x_{0}$ is $a_{1g}$, and  $x_{1}$ and $x_{2}$ are of $e_{g}^{\pi}$ symmetry (see Fig.~2 of the main text for a crystal field diagram). The trigonal basis is related to the cubic one via the basis transformation ($u_{+}$,$u_{-}$,$x_{0}$, $x_{1}$, $x_{2}$)=B$_{c\rightarrow t} (xy, yz, 3z^2-r^2, xz, x^2-y^2)$~\cite{sugano_multiplets_1970}, where 

\begin{equation}
B_{c\rightarrow t} =\frac{1}{\sqrt[]{6}}
\left(\begin{array}{ccccc}

i & -\sqrt[]{2} i & 0 & -\sqrt[]{2} & -1\\
i & -\sqrt[]{2} i & 0 & \sqrt[]{2} & 1\\
0 & 0 & \sqrt[]{6} & 0 & 0\\
\sqrt[]{2} i & i & 0 & 1 & -\sqrt[]{2}\\
\sqrt[]{2}  i& i & 0 & -1 & \sqrt[]{2}
\end{array}\right).
\end{equation}

The coordinate transformation between the bases is then given by $T_{c\rightarrow t}=(B_{c\rightarrow t}^{-1})^T$, and the crystal field Hamiltonian in the cubic basis can be found as $H_{CF}^{c}=T_{c\rightarrow t}^{-1}H_{CF}^{t}T_{c\rightarrow t}$, where $H_{CF}^{t}$ is the diagonal crystal field Hamiltonian in the trigonal basis. Allowing for both the octahedral (C$_{o}$) and trigonal (C$_{t}$) crystal field, $H_{CF}^{t}$ is given by 

\begin{equation}
H_{CF}^{t}=
\left(\begin{array}{ccccc}

C_{o} & 0 & 0 & 0 & 0\\
0 & C_{o} & 0 & 0 & 0\\
0 & 0 & C_{t} & 0 & 0\\
0 & 0 & 0 & 0 & 0\\
0 & 0 & 0 & 0 & 0
\end{array}\right).
\end{equation}

In the simplified model I (Extended Data Fig.~3) the $e_{g}^{\sigma}$ orbitals are completely neglected, which is technically achieved by applying a very large octahedral crystal field. An overall on-site energy, E$_{Co}$ is added to all of the Co states.  The on-site energy of the in-plane and out-of-plane $2p$ orbitals of the first (second) oxygen are labelled as $E_{1}^{xy}$ and $E_{1}^{z}$ ($E_{2}^{xy}$ and $E_{2}^{z}$), respectively. In the simplified model I, which considers only the $p_{1,2}^{z}$ orbitals, large on-site energies $E_{1,2}^{xy}$ are applied. 

The spin-orbit Hamiltonian is given by, $H_{SO}=\xi \bf{L} \cdot \bf{S}$, where $\xi$ is the atomic spin-orbit coupling constant of the Co $3d$ orbitals, taken to be equal to $70meV$~\cite{haverkort_spin_2005}. The Hamiltonian in the cubic basis $(xy\downarrow, yz\downarrow, 3z^2-r^2\downarrow, xz\downarrow, x^2-y^2\downarrow, xy\uparrow, yz\uparrow, 3z^2-r^2\uparrow, xz\uparrow, x^2-y^2\uparrow)$, where $\uparrow$ ($\downarrow$) stand for spin up (down), is given by 

\begin{equation}
H_{SO}=\frac{\xi}{2}
\left(\begin{array}{cccccccccc}

0 & 0 & 0 & 0 & -2i & 0 & -1 & 0 & -i & 0\\
0 & 0 & 0 & -i & 0 & 1 & 0 & -\sqrt[]{3} & 0 & -i\\
0 & 0 & 0 & 0 & 0 & 0 & \sqrt[]{3} i  & 0 & \sqrt[]{3} & 0\\
0 & i & 0 & 0 & 0 & i & 0 & -\sqrt[]{3} & 0 & 1\\
2i & 0 & 0 & 0 & 0 & 0 & i & 0 & -1 & 0\\
0 & 1 & 0 & -i & 0 & 0 & 0 & 0 & 0 & 2i\\
-1 & 0 & -\sqrt[]{3} i & 0 & -i & 0 & 0 & 0 & i & 0\\
0 &  -\sqrt[]{3} & 0 & \sqrt[]{3} & 0 & 0 & 0 & 0 & 0 & 0\\
i & 0 & \sqrt[]{3} & 0 & -1 & 0 & -i & 0 & 0 & 0\\
0 & i & 0 & 1 & 0 & -2 i & 0 & 0 & 0 & 0\\
\end{array}\right).
\end{equation}

The two sets of parameters used to calculate the tight-binding band structure are shown in Extended Data Table~3.  Model I is maximally simplified, retaining only Co-O nearest-neighbour hopping between Co t$_{2g}$ and O p$_z$ orbitals as discussed above, in order to isolate the  key ingredients necessary for the large inversion symmetry breaking energy scale (Extended Data Fig.~3). Co-O hopping is parametrised via two Slater-Koster parameters, $V_{dp\sigma}$ and $V_{dp\pi}$, employing the empirical relation  $V_{dp\pi}=-\sqrt[]{3}/4V_{dp\sigma}$~\cite{wu_orbital_2005}. The on-site energies of the oxygen $p_{z}$ orbitals are estimated from the DOS shown in Extended Data Fig.~2, and chosen to be $E_{O_2}=-7eV$ ($E_{O_1}=-3.2eV$) for oxygen below (above) Co. Guided by the bandwidth and filling of the CoO$_2$-derived surface states from our DFT supercell calculations,   the total bandwidth is required to be $\sim 500meV$, and the Fermi level to be $\sim 100 meV$ below the top of the band. This sets the on site energy of the Co orbitals to be  $-600meV$, and the hopping parameter $V_{dp\sigma}$ to 1.2~eV, a value similar to those obtained from fits to the DFT band structure of the related compound Na$_{x}$CoO$_{2}$ \cite{johannes_tight-binding_2004}. The differences between the  oxygen and cobalt on-site energies, relevant for determining effective hopping parameters, are $\Delta_{O_1}=-2.6eV$ and $\Delta_{O_2}=-6.4eV$.  For the calculation with no asymmetry (Extended Data Fig.~3) the on-site energy difference was required to satisfy  $2/\Delta_{O}=1/\Delta_{O_1}+1/\Delta_{O_2}$, yielding the oxygen on-site energy of $E_{O}=-4.3eV$.

Model II additionally incorporates direct Co - Co and O - O hopping, as well as the full Co $3d$ and O $2p$ orbital manifolds.
As the in-plane oxygen orbitals do not directly participate in the Pt - O bonding, their on-site energy is allowed to be different to that of the $p_{z}$ orbitals. In particular, the DFT PDOS suggests the difference between $E_{1}^{x,y}$ and $E_{2}^{x,y}$ is $\sim2~eV$ smaller than that between $E_{1}^{z}$ and $E_{2}^{z}$. 
Allowing these different hopping paths, plausible for the real material, increases the degree of orbital mixing, and thus spin-splitting, across $k$ space. The calculated electronic structure for this model is shown in Extended Data Fig.~7.

}

\

\noindent{\bf Acknowledgements}\\
{\small
\noindent We thank Nabhanila Nandi and Burkhard Schmidt for useful discussions. We gratefully acknowledge support from the European Research Council (through the QUESTDO project), the Engineering and Physical Sciences Research Council, UK (Grant No.~EP/I031014/1), the Royal Society, the Max-Planck Society and the International Max-Planck Partnership for Measurement and Observation at the Quantum Limit. VS, LB, OJC and JMR acknowledge EPSRC for PhD studentship support through grant Nos.~EP/L015110/1, EP/G03673X/1, EP/K503162/1, and EP/L505079/1. DK acknowledges funding by the DFG within FOR 1346. We thank Diamond Light Source and Elettra synchrotrons for access to Beamlines I05 (Proposal Nos.~SI12469 and SI14927) and APE (Proposal No.~20150019), respectively, that contributed to the results presented here.

\newcommand{\beginsupplement}{%
        \setcounter{table}{0}
        \renewcommand\tablename{Extended Data Table}
        \renewcommand{\thetable}{\arabic{table}}%
        \setcounter{figure}{0}
        \renewcommand\figurename{Extended Data Figure}
     }
     
     \beginsupplement

%%parameter table
\begin {table*}[h]
\begin{center}
    \begin{tabular}{ |c | c | c| c|}
    \hline
    	  &  PtCoO$_{2}$ & PdCoO$_{2}$ & PdRhO$_{2}$ \\ \hline \hline

			$m^*_1$/  $m_{e}$ [$\overline{\Gamma}$-$\overline{M}$] & 9.6 $\pm$ 0.5  & 7.9 $\pm$ 0.5 & 6 $\pm$ 1 \\ \hline
            $m^*_2$/  $m_{e}$ [$\overline{\Gamma}$-$\overline{M}$] & 11.5 $\pm$ 0.8 & 10.1 $\pm$ 0.7 &  7 $\pm$ 1 \\ \hline

			$m^*_1$/  $m_{e}$ [$\overline{\Gamma}$-$\overline{K}$] & 9.5 $\pm$ 0.5 & 9.0 $\pm$ 0.4 & 6.0 $\pm$ 0.5 \\ \hline
            $m^*_2$/ $m_{e}$ [$\overline{\Gamma}$-$\overline{K}$] & 15 $\pm$ 1 & 16 $\pm$ 1 & 11 $\pm$ 1 \\ \hline\hline

             $\Delta k_{F}$/ \AA $^{-1}$ [$\overline{\Gamma}$-$\overline{M}$] & 0.11 $\pm$ 0.01 & 0.09 $\pm$ 0.01 & 0.13 $\pm$ 0.01 \\ \hline
            $\Delta k_{F}$/ \AA $^{-1}$ [$\overline{\Gamma}$-$\overline{K}$] & 0.13 $\pm$ 0.01 & 0.12 $\pm$ 0.01 & 0.16 $\pm$ 0.01\\ \hline\hline
             $\Delta E$/ meV [$\overline{K}$] &  & 60 & 150 \\ \hline

\end{tabular}

\caption{ \label{t:CompoundsFermi} {\bf Comparison of the spin-split surface states of delafossite oxides} Quasiparticle masses, m$^*_{i}$, of the inner ($i=1$) and outer ($i=2$) surface bands, and spin splitting at the Fermi level, $\Delta k_{F}$, along the high symmetry directions.  These are very similar for PtCoO$_{2}$ and PdCoO$_{2}$. Despite the lower masses for PdRhO$_{2}$, $\Delta k_{F}$ is larger, a consequence of the larger energetic splitting $\Delta E$.}

\end{center}
\end{table*}

%%tight binding table
\begin {table*}[b]
\begin{center}
    \begin{tabular}{ |c | c | c|c|}
    \hline

	Hopping path & $\theta$ & $\phi$ & SK parameters \\ \hline
    Co-Co & $90$ & $30$, $90$, $150$, $210$, $270$, $330$  & $V_{dd\sigma}$, $V_{dd\pi}$, $V_{dd\delta}$ \\ \hline
     Co-O1 & $60.08$ & 0, $120$, $240$ & $V_{dp\sigma}$, $V_{dp\pi}$\\ \hline
      Co-O2 & $119.92$ &  $60$, $180$, $300$ & $V_{dp\sigma}$, $V_{dp\pi}$ \\ \hline
      O1-O1 (O2-O2) & $90$ & $30$, $90$, $150$, $210$, $270$, $330$  & $V_{pp\sigma}$, $V_{pp\pi}$\\ \hline
      O1-O2  & $139$ & $30$,$150$, $270$   & $V_{pp\sigma}$, $V_{pp\pi}$\\ \hline

\end{tabular}

\caption{ \label{t:SK} {\bf Tight binding model.} The hopping paths considered in the tight binding model, along with angles (expressed in degrees) between nearest neighbour atoms, and relevant Slater-Koster parameters. }

\end{center}
\end{table*}

%%parameter table
\begin {table*}[h]
\begin{center}
    \begin{tabular}{ |c | c | c|}
    \hline
    
  	Parameter & Model I & Model II\\ \hline
    $V_{dd\sigma}$ & 0  & -0.3\\ \hline
    $V_{dd\pi}$ & 0 & 0.15\\ \hline
    $V_{dd\delta}$  & 0  & 0\\ \hline
    $V_{dp\sigma}$  & -1.2 & -1.6\\ \hline
    $V_{dp\pi}$  & 0.52  & 0.87\\ \hline 						$V_{pp\sigma}$  & 0 &  0.5\\ \hline
	$V_{pp\pi}$  &0 & -0.3\\ \hline
    $C_{o}$  & 1000 & 1\\ \hline
	$C_{t}$  & 0 & 0.5\\ \hline
    $E_{1}^{z}$  & -3.2 & -3.2\\ \hline
    $E_{2}^{z}$   & -7 & -7\\ \hline
    $E_{1}^{x,y}$ & 1000 & -4\\ \hline
    $E_{2}^{x,y}$ & 1000 & -6\\ \hline
    $E_{Co}$ & -0.57 & -1.2\\ \hline
    $\xi$ & 0.07 & 0.07\\ \hline
   
\end{tabular}

\caption{ \label{t:param} {\bf Tight-binding parametrisation.} The  parameters used in the two versions of the model (Extended Data Figs.~4 and 7 for Model I and Model II, respectively), all expressed in eV. The meaning of the symbols is explained in the Methods section. The large values of 1000~eV for on-site energies are used to effectively remove corresponding states from the model.}

\end{center}
\end{table*}

\begin{figure*}[h]
\begin{center}
\includegraphics[width=0.8\textwidth]{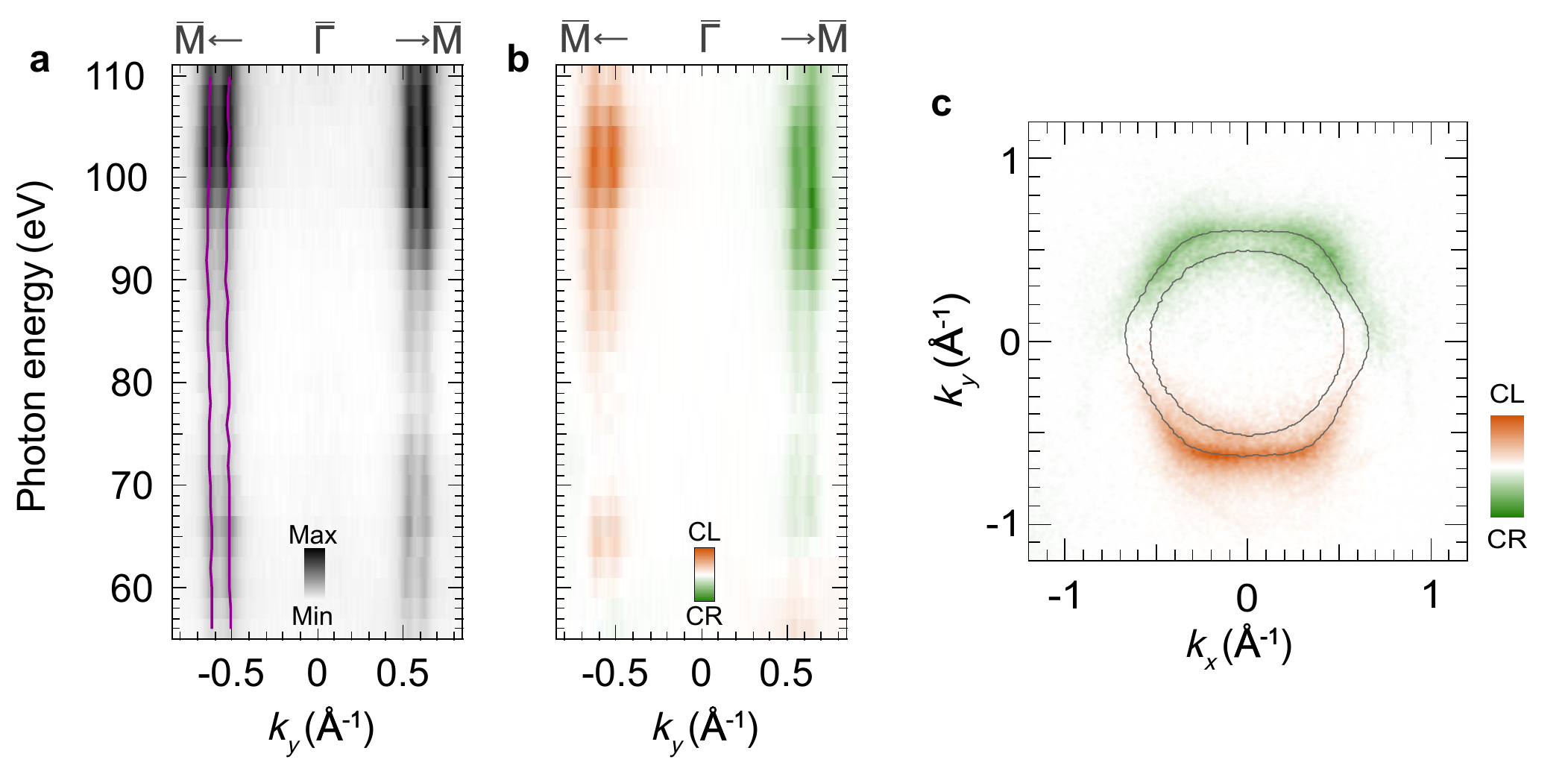}
 \caption{ \label{f:hvu} {\bf Photon energy and polarisation dependent ARPES of PtCoO$_{2}$.} (a)  The intensity at the Fermi level (E$_{F}\pm$5~meV, sum of measurements with circularly left (CL) and right (CR) polarised light), as a function of incident photon energy. The full purple lines correspond to the peak positions of fits to momentum distribution curves at the Fermi level. The Fermi crossing vectors do not depend on the photon energy, indicating that the states attributed to the CoO$_{2}$ surface layer are indeed two dimensional. (b) A strong circular dichroism of these states is evident over an extended photon energy range, and (c) with an in-plane momentum dependence indicating the same chirality of orbital angular momentum for the two spin split bands (h$\nu$ =110~eV). The grey lines in (c) represent the Fermi momenta extracted from the sum of the measurements in CL and CR polarisations by fitting momentum distribution curves (MDCs) radially around the Fermi surface.}
\end{center}
\end{figure*}

\begin{figure*}[h]
\begin{center}
\includegraphics[width=0.7\textwidth]{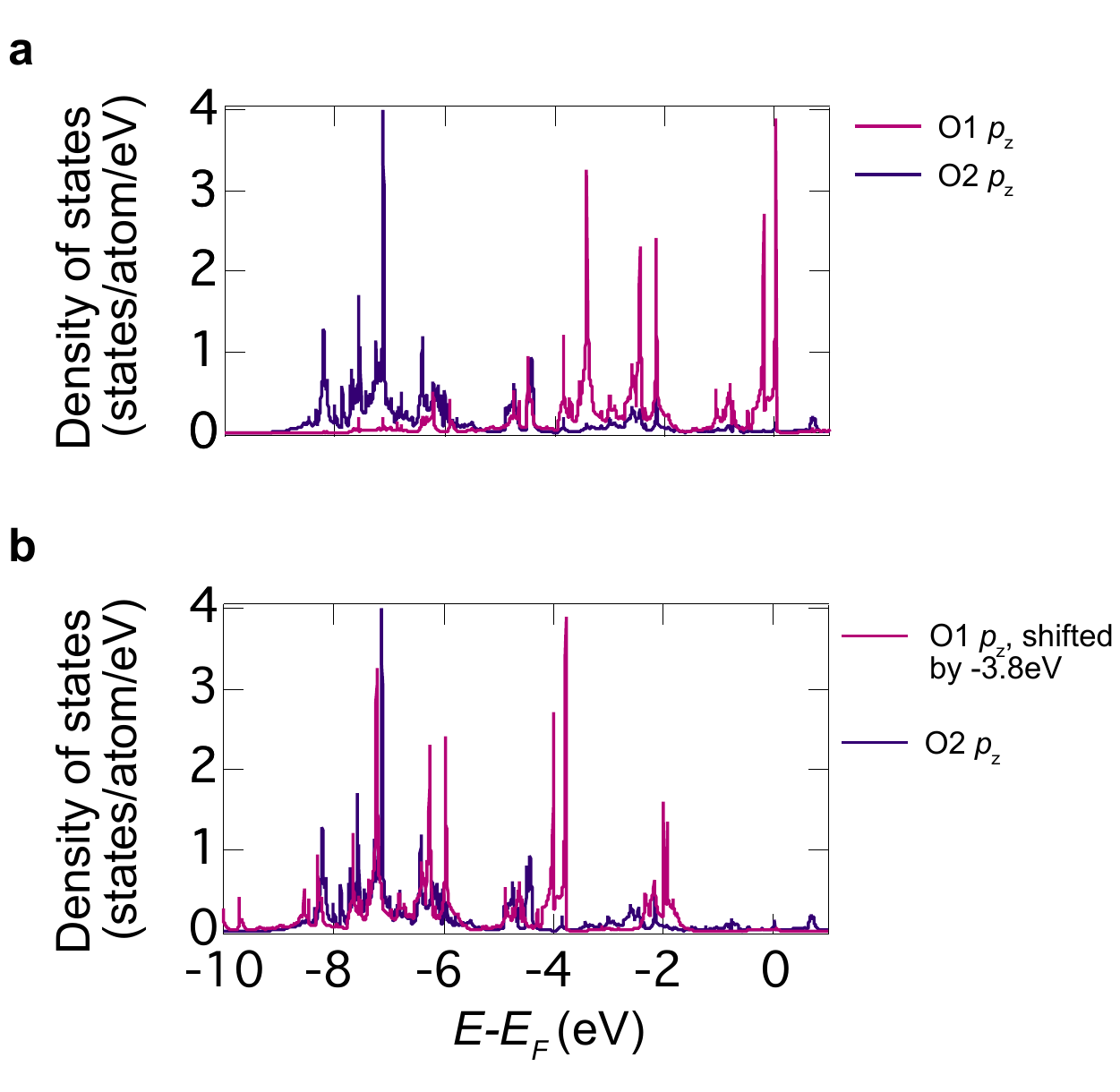}
\caption{ \label{f:O} {\bf Partial density of states of oxygen in the first CoO$_{2}$ layer.} (a) The oxygen p$_z$ partial density of states (PDOS) for layers above (O1, pink) and below (O2, purple) the Co layer. The O1 p$_{z}$  PDOS is much larger than that of O2 close to the Fermi level (see also Fig. 4(c,d) of the main text). (b) Except for this added weight, the O1 p$_{z}$ PDOS shifted by  -3.8 eV in binding energy well approximates that of O2.} 
\end{center}
\end{figure*}

\begin{figure*}[!t]
\begin{center}
\includegraphics[width=0.7 \textwidth]{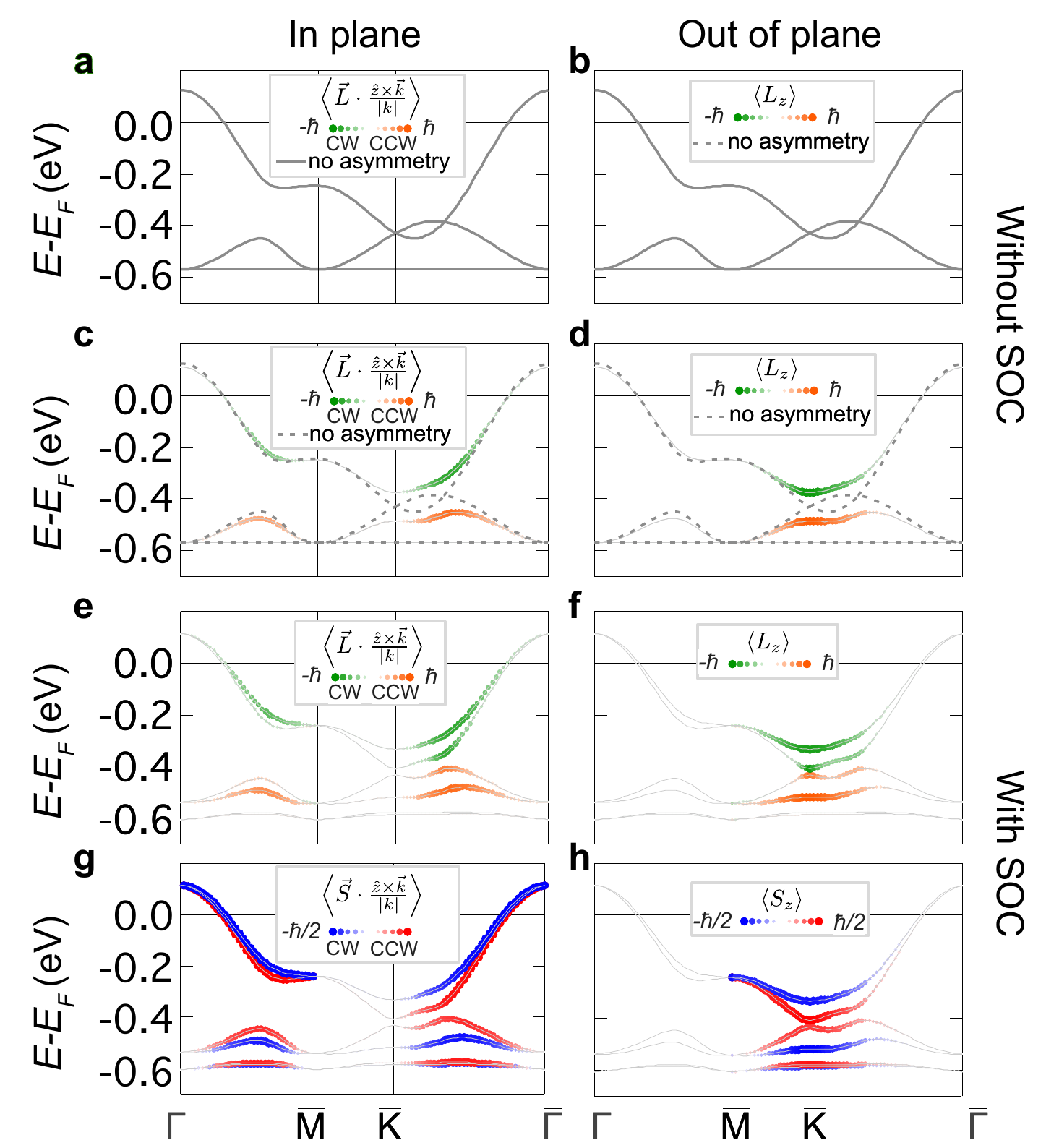}
\caption{ \label{f:TB} {\bf Development of orbital and spin angular momentum.} Band structure obtained from a minimal tight-binding model (see Methods) revealing the key ingredients to develop maximal Rashba-like spin splitting. The calculations are shown (a,b) without SOC or ISB, (c,d) with ISB but no SOC, and (e-h) with ISB and SOC. The chiral clockwise (CW) and counter-clockwise (CCW) in-plane (c,e) and out-of-plane (d,f) orbital angular momentum and chiral CW/CCW in-plane (g) and out-of-plane (h) spin angular momentum are shown by colouring (see legends). If the two oxygens have the same on-site energy (no asymmetry, $E_{O1}$=$E_{O2}$), and neglecting spin-orbit coupling (a,b), the electronic structure closely resembles that of a Kagome model, which has been used previously to describe the CoO$_{2}$ layer of Na$_{x}$CoO$_{2}$~\cite{koshibae_electronic_2003} -- the lowest band is flat, and the other two bands cross at the Brillouin zone corner, $\overline{K}$, and along the $\overline{\Gamma}-\overline{K}$ line, where hybridisation is forbidden by symmetry. Orbital angular momentum is quenched in this inversion-symmetric environment. (c,d) Introducing asymmetry due to a difference in the on-site energy of O1 and O2 allows orbital mixing, and hybridisation gaps open where there are crossings in the absence of the symmetry breaking. The orbital mixing allows these bands to develop a large  (magnitude approaching $\hbar$) orbital angular momentum (OAM) even in the absence of spin-orbit coupling. This is largely chiral (OAM perpendicular to in-plane momentum) along the $\overline{\Gamma}-\overline{K}$ and $\overline{\Gamma}-\overline{M}$ directions, and crosses over to  the OAM having a significant out-of-plane component close to $\overline{K}$ where any in-plane component must vanish due to symmetry. For such an in-plane and out-of-plane OAM to develop there must be an out-of-plane and in-plane inversion symmetry breaking (ISB), respectively. The fact that the asymmetric hopping occurs via the layers above and below Co naturally gives rise to the out-of-plane ISB. The opposite orientation of nearest-neighbour Co-O bonds to the oxygen layers above and below the transition metal plane (Fig.~\ref{f:DFT}(b)) additionally provides the in-plane ISB. Together, this allows the OAM to remain large across a greater portion of the Brillouin zone, rather than being suppressed in the broad vicinity of the $\overline{K}$ point. Crucially, the hybridisation gaps between states of opposite orbital angular momentum opened by such inversion symmetry breaking are as large as 140 meV  for the realistic parameters used here, which is about twice the size of atomic SOC. This difference in energy scales means the spin-orbit interaction introduces an additional splitting between the states of spin parallel and anti-parallel to the pre-existing OAM, which is itself not significantly altered (e-h). The energetic splitting assumes the full value of the atomic SOC, validating the simple schematic shown in Fig.~\ref{f:overview}(b) of the main text. This picture is consistent with both our spin-resolved ARPES (Fig.~\ref{f:spin}(b)) and our CD-ARPES (Extended Data Fig.~1 and Fig.~4), which show that the two spin-split branches of the CoO$_2$-derived surface states host the same sign of OAM.}
\end{center}
\end{figure*}

\begin{figure*}[!t]
\begin{center}
\includegraphics[width=\textwidth]{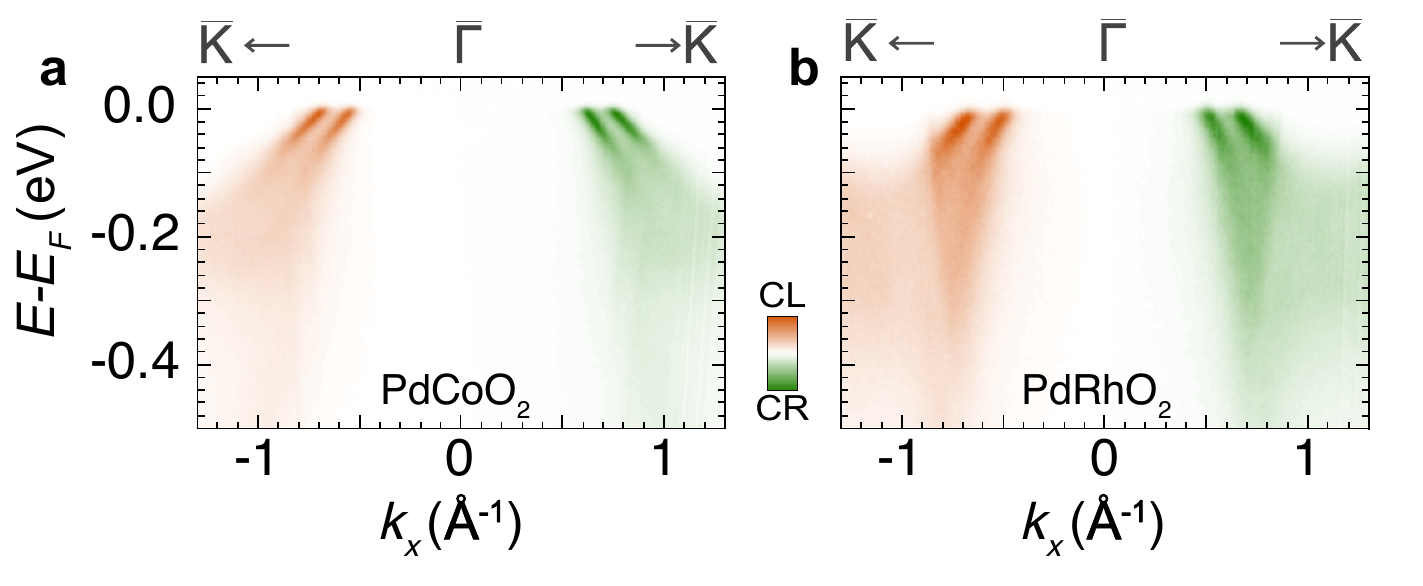}
\caption{ \label{f:CD} {\bf Circular dichroism ARPES of PdCoO$_2$ and PdRhO$_2$. }   Similar to PtCoO$_2$ (Fig. \ref{f:PtCoO2_summary} and Extended Data Fig.~1)), PdCoO$_2$ and PdRhO$_2$ surface states  show strong circular dichroism (h$\nu$ =90~eV), of the same sign for each of the two spin-split branches.  This is  consistent with the results of our minimal tight binding  model (Extended Data Fig.~3), and confirms how the large inversion symmetry breaking leads to chiral orbital angular momentum for transition metal derived surface states across the delafossite oxide family . 
}
\end{center}
\end{figure*}

 \clearpage

\begin{figure*}[!t]
\begin{center}
\includegraphics[width=0.68 \textwidth]{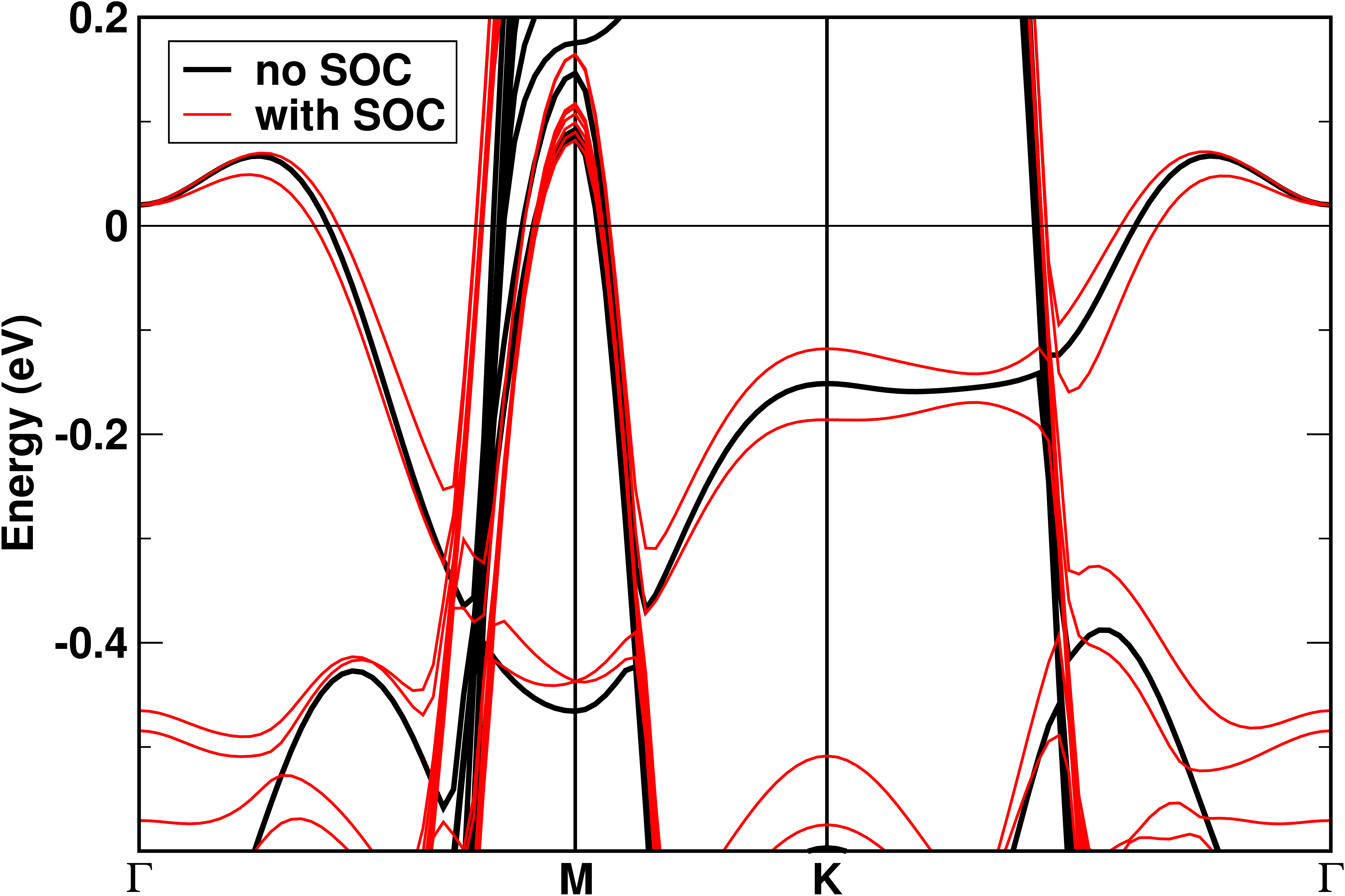}
\caption{ \label{f:bandsSOC} {\bf The influence of spin-orbit coupling on the band structure.} Zoom of the band structure of PtCoO$_2$ around the Fermi level. For the narrow Co-O surface band (between about -0.45 and 0.1 eV), the spin orbit coupling (SOC, red lines) leads to a spin splitting of this band, with only small changes to the dispersion. 
}
\end{center}
\end{figure*}

\begin{figure*}[!t]
\begin{center}
\includegraphics[width=0.68 \textwidth]{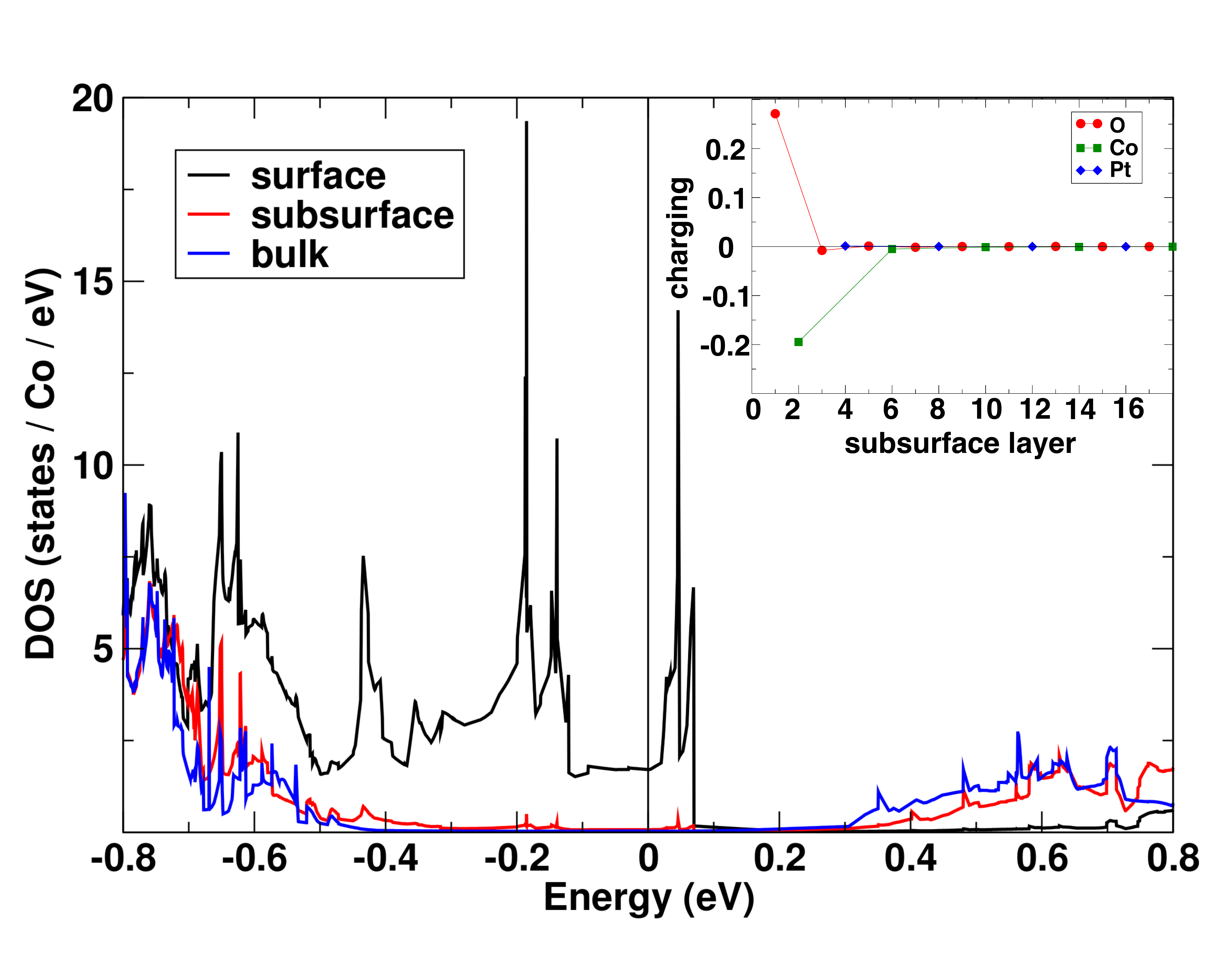}
\caption{\label{f:DOScharge} {\bf Subsurface Co contribution.} 
Calculated Co partial density of states near the Fermi level for different Co layers. The surface state between $\sim$ -0.45 and $\sim$ 0.1 eV (see Fig. \ref{f:bandsSOC}) has very little contribution from Co atoms below the first layer. The partial density of states of the subsurface Co atoms is almost bulk-like. This is also reflected by the charging of the surface shown in the inset. The plot shows the additionally accumulated charge vs depth below the surface, referenced to the constituent bulk charges.  Only the surface O and the topmost Co-layer deviate significantly from the bulk. In particular, the pronounced difference (asymmetry) between the two O of the CoO$_2$ surface layer is very clearly demonstrated. Surface relaxation has only a minor impact to this scenario.
}
\end{center}
\end{figure*}

\begin{figure*}[h]
\begin{center}
\includegraphics[width=0.8\textwidth]{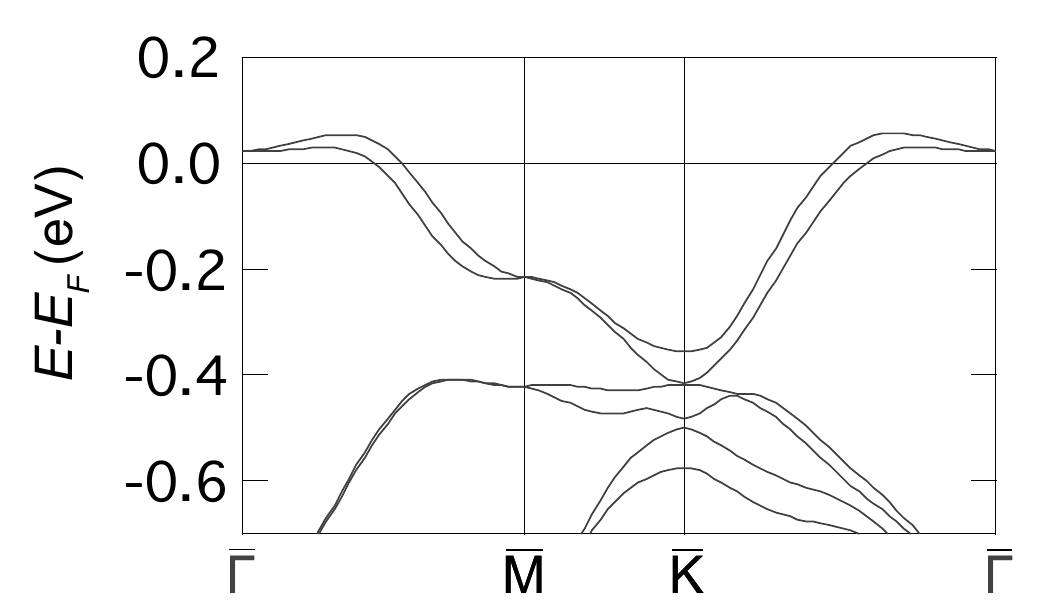}
\caption{ \label{f:TB_2} {\bf Tight binding model II.} The band structure calculated using the tight-binding model II (see Table \ref{t:param}). Additional hopping paths allowed in this model increase the orbital mixing, and thus the spin splitting, across the $k$-space.}
\end{center}
\end{figure*}

\end{document}